\documentclass[runningheads]{llncs}
\usepackage{graphicx}
\usepackage{amsmath}
\usepackage{multicol}
\usepackage{csquotes}
\usepackage{amsfonts}
\usepackage{cite}
\usepackage{submission}
\usepackage{macros}

\setmode{final}

\begin{document}

\title{Real bird dataset with imprecise and uncertain values}
\titlerunning{Real bird data set with imprecise and uncertain values}

\author{Constance Thierry \and Arthur Hoarau \and
Arnaud Martin \and
Jean-Christophe Dubois \and 
Yolande Le Gall}

\authorrunning{C. Thierry et al.}
\institute{Univ Rennes, CNRS, IRISA, DRUID, France}

\maketitle           


\begin{abstract}
The theory of belief functions allows the fusion of imperfect data from different sources. Unfortunately, few real, imprecise and uncertain datasets exist to test approaches using belief functions. 
We have built real birds datasets thanks to the collection of numerous human contributions that we make available to the scientific community. 
The interest of our datasets is that they are made of human contributions, thus the information is therefore naturally uncertain and imprecise. These imperfections are given directly by the persons. 
This article presents the data and their collection through crowdsourcing and how to obtain belief functions from the data. 

\keywords{Datasets \and Imprecise \and Uncertain}
\end{abstract}


\section{Introduction} \label{section:introduction}

The theory of belief functions allows for uncertainty and imprecision in the data. However, there are very few real datasets available to consider belief functions.
\cite{dubois19} and \cite{abassi18} work on imprecise and uncertain real data that the authors have collected but these data are not made available to the community.
Similarly, \cite{diaz08} proposes an MCQ that allows students to give imprecise and uncertain answers that can be modeled with belief functions. However, the experimental data are not reported. 
It was important to build real datasets to evaluate proposed methods in real context~\cite{thierry2021}. 
To do so, we collected human contributions from crowdsourcing campaigns, as human contributions are uncertain and imprecise information.

Crowdsourcing is the outsourcing of tasks to a crowd of contributors on a platform dedicated to the domain \cite{howe06}. The tasks that can be achieved through crowdsourcing are very diverse. 
In this paper, we presented to the contributors a picture of a bird and asked them to identify the bird from a list of proposed names.
We use interfaces that allow us to collect imprecise and/or uncertain responses.
We conducted six crowdsourcing campaigns for bird photo annotation. 
For all these campaigns the contributor had to give his certainty in his answer. 
Two of them are only composed of precise contributions. 
For the four other campaigns, the contributor can be imprecise and choose more bird names. 
For two of the imprecise campaigns, after the contributor has given his answer he is offered to enlarge or restrict his selection consequently.

The rest of the paper is as follows, section \ref{section:BF} introduces the belief functions. 
Section \ref{section:Campaign} reviews the crowdsourcing campaigns and section \ref{section:data} presents the datasets.
We propose examples of modelisation thanks to the belief function section \ref{section:model}.
Section \ref{section:conclusion} concludes the paper.

\section{Belief functions} \label{section:BF}

The theory of belief functions, also called Dempster-Shafer theory~\cite{Dempster1967, shafer1976}, is used in this study in order to model both data imprecision and uncertainty.

One considers $\Omega = \{r_1, \ldots, r_M\}$ the frame of discernment for $M$ exclusive and exhaustive hypotheses. In this paper, $\Omega$ represents all possible bird species of a given photo among $M$ bird species.
The power set $2^\Omega$ is the set of all subsets of $\Omega$.
A basic belief assignment is the belief that a source may have about the elements of the power set of $\Omega$, this function assigns a mass to each element of this power set such that the sum of all masses is equal to $1$.

\begin{equation} \label{eq:unity}
    \begin{split}
    m : 2^\Omega\rightarrow [0, 1]\\
    \sum_{A\in 2^\Omega} m(A) = 1 
    \end{split}
\end{equation}

\begin{description}
    \item [Focal element:]An element of $2^\Omega$ with a non-null mass.
    \item [Simple support mass function:]Only has two focal elements, and one of them is the frame of discernment $\Omega$.
    \item [Consonant mass function:] Each focal element is nested.
\end{description}

\section{Crowdsourcing Campaign} \label{section:Campaign}

The main objective for these campaigns is always the same, a photo of a bird is presented to the contributor with a set of species names (including the good answer) and he has to select the right answer.
The Wirk platform (Crowdpanel\footnote{https://crowdpanel.io/ (15/04/2022)}) is used to realize the crowdsourcing campaigns.
As the users of the platform live in France, the birds used for the campaigns are all of species visible in metropolitan France.
For all the campaigns, the contributor has to give his answer, then specify his certainty according to the following Likert scale:
``Totally uncertain'',``Uncertain'', ``Rather uncertain'', ``Neutral'', ``Rather certain'', ``Certain'', ``Totally certain''.
We explained to the contributors that there is no penalty for being uncertain and/or imprecise in their answers.
After having given his answer and his certainty, he can validate his contribution in order to move on to the next question.

\subsection{Campaigns multi\_birds}

For these two campaigns, five bird names are proposed to the contributor. 
The names change from one question to another and a bird species is presented only once.
We have tried to introduce different levels of difficulty in the questions.
For example, for a difficult question, a photo of an eagle is presented to the contributor and the five answer items are different species of eagles.
Conversely, for a simpler question, a photo of a gull is presented to the contributor and the four other answers are names of duck species.
For a single photo, responses were presented in random order to each contributor to avoid selection bias. 
In addition, the questions were also asked in a random order, so that when a contributor $c_1$ answers a question $q_i$, $c_2$ answers $q_j$.
These crowdsourcing campaigns include 3 attention questions for which the contributor is asked to give the same answer as the one given in the previous question.

\begin{figure}[t]
\begin{tabular}{cc}
     \includegraphics[height=4.6cm]{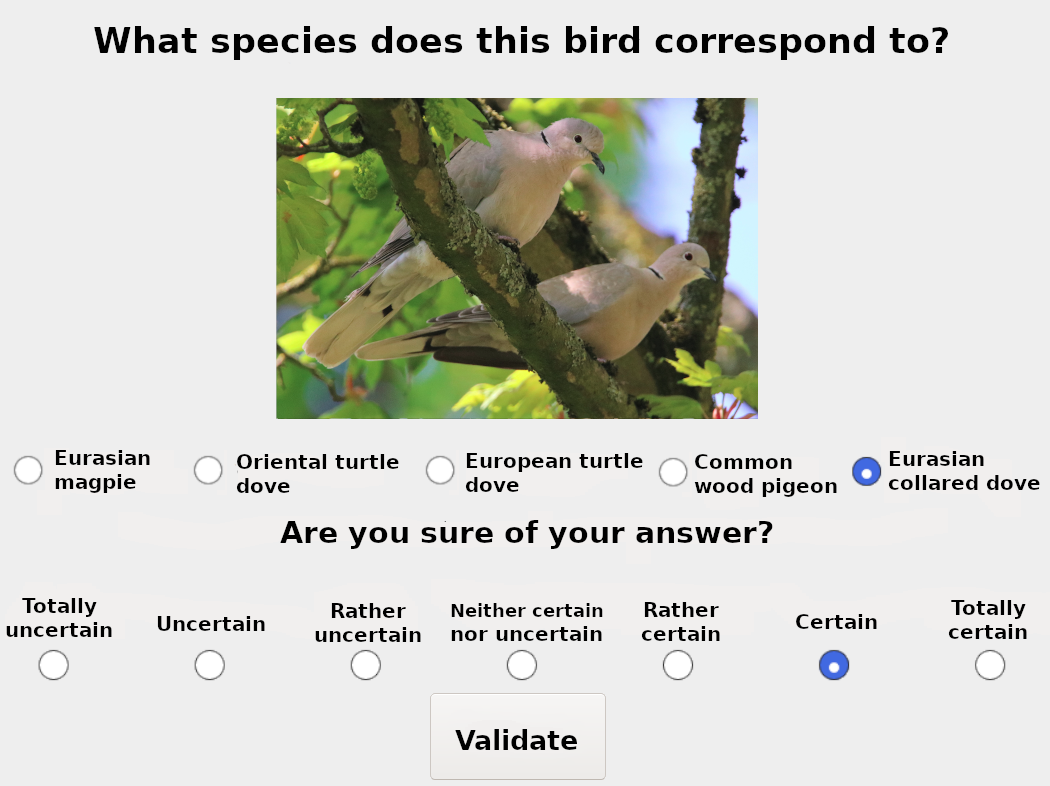} \quad & \includegraphics[height=4.6cm]{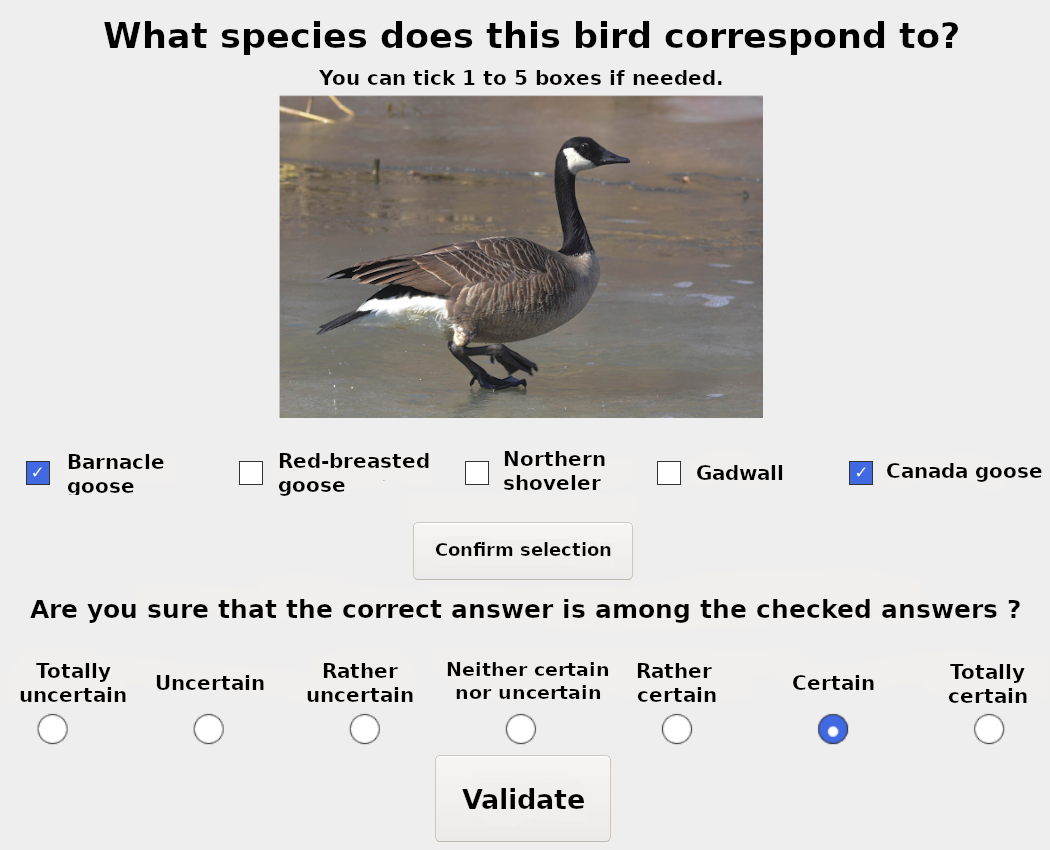} \\
    (a) multi\_birds\_precise & (b) multi\_birds\_imprecise
\end{tabular}
    \caption{Interfaces used for crowdsourcing campaigns multi\_birds}
    \label{fig:interfaces_multi}
\end{figure}

{\bf multi\_birds\_precise}
The interface used for this task is given figure \ref{fig:interfaces_multi}.(a).
Participants have to provide a precise answer by selecting a single bird name, and a self-assessment of their certainty in this answer.


{\bf multi\_birds\_imprecise}
For this task the contributor can be imprecise and select up to all of the bird names offered. The interface is given figure \ref{fig:interfaces_multi}.(b).
The contributors first must give his answers, validate it and then he is asked to give his certainty in this answer.

For both campaigns the crowds are composed of 100 contributors, each one must annotate 50 photos, for a total of 5000 contributions for each campaign.
A contributor allowed to do the first campaign cannot participate in the second.

\subsection{Campaigns 10\_birds}

For these campaigns, ten bird species are selected and proposed as response elements to the contributors.
In order to observe the contributor's ability to be imprecise in case of hesitation, the ten birds presented are composed of subgroups from the same bird family given table \ref{tab:oiseaux_10}.
\begin{table}[t]
    \centering
    \caption{The ten bird species used in the 10\_birds campaigns group by family}
    \begin{tabular}{|l|l|l|l|}
        \hline
        Muscicapidae & Columbidae & Paridae & Corvidae\\
        \hline
        European robin & Common wood pigeon & Great tit & European jackdaw \\
        & Rock pigeon & Marsh tit & Carrion crow\\
        &             & Coal tit  & Common raven\\
        &&& Rook crow \\
    \hline
    \end{tabular}
    \label{tab:oiseaux_10}
\end{table}
The bird names are presented to each contributor in a different order to avoid selection bias.
This ordering of names is nevertheless fixed for a contributor throughout the campaign.
Such as the campaigns multi\_birds, the  questions are asked in a random order. 
The same scale is used for certainty and 3 attention questions are also asked.
The contributor is no longer required to validate his answer before he can give his certainty.

\begin{figure}[t]
\begin{tabular}{cc}
     \includegraphics[height=4.2cm]{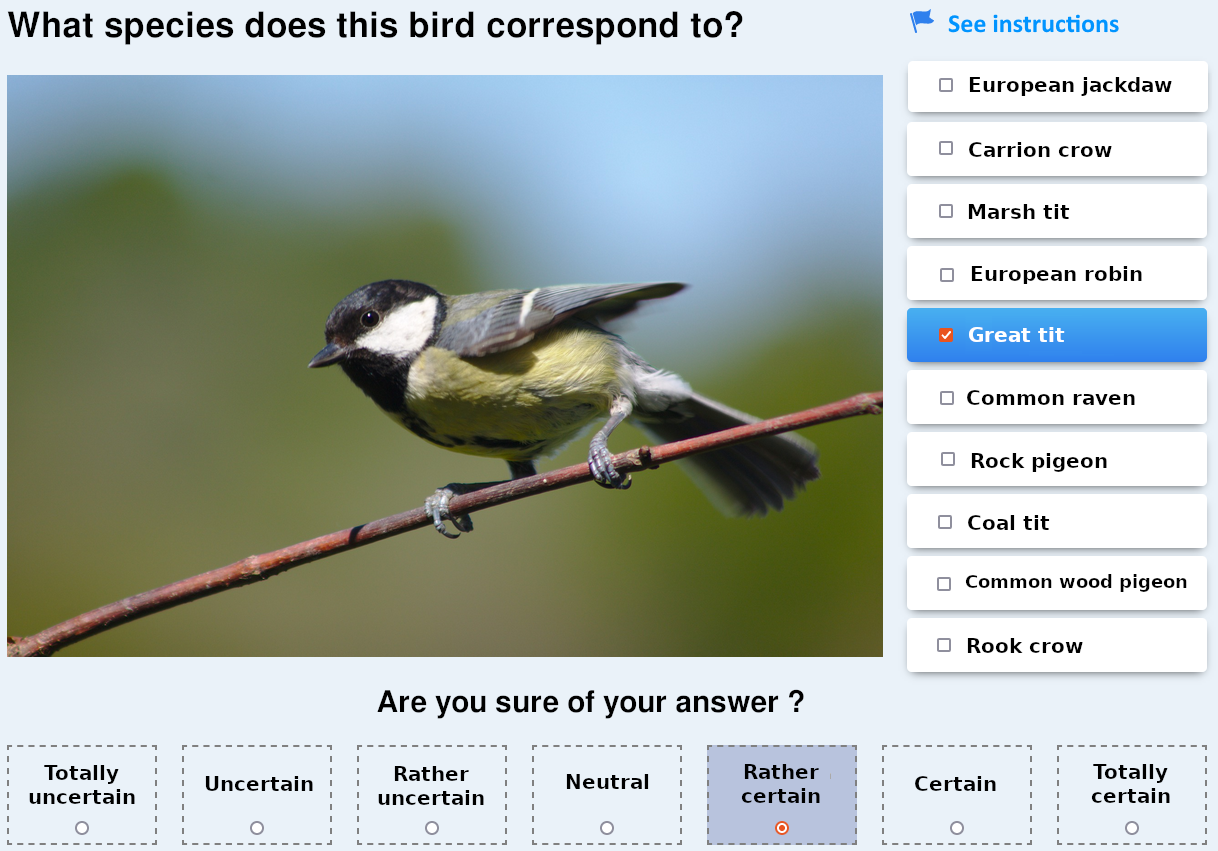} \quad & \includegraphics[height=4.2cm]{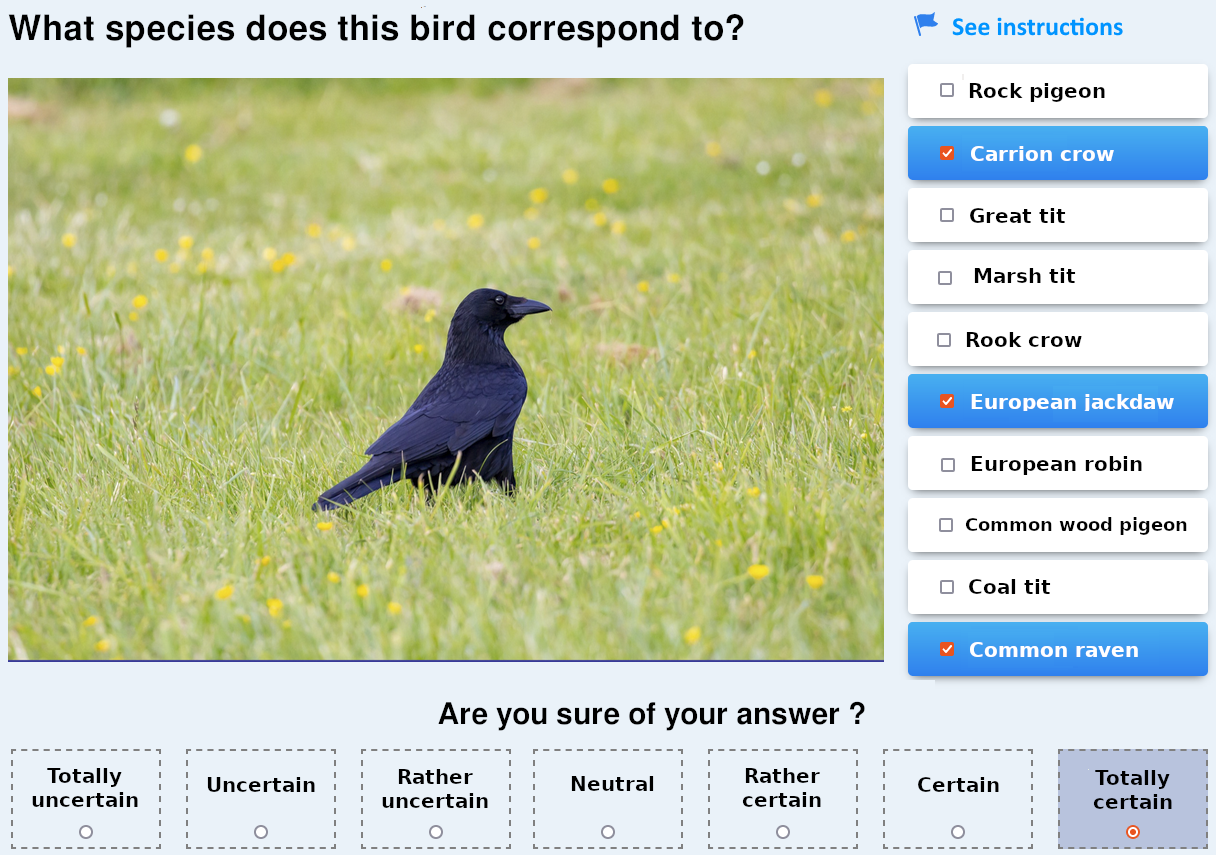} \\
    (a) precise & (b) imprecise
\end{tabular}
    \caption{Interfaces used for crowdsourcing campaigns 10\_birds precise and imprecise}
    \label{fig:interfaces_10}
\end{figure}


{\bf 10\_birds\_precise}
The contributor should select from the interface figure \ref{fig:interfaces_10}(a) a unique bird name and then give his certainty about it.


{\bf 10\_birds\_imprecise}
The contributor can choose thanks to the interface figure \ref{fig:interfaces_10}(b) 1 to a maximum of 5 answers from the ten provided bird names. 
We impose a maximum number of answers to 5 because we admit that if the contributor hesitates it is between names of birds of the same family. He should not hesitate between a pigeon and a chickadee for example. 
We have chosen to offer the crowd a maximum selection of 5 names because we do not want to introduce a bias and encourage him to choose exactly the 4 corvidaes in case of hesitation.

\begin{figure}[t]
\begin{tabular}{cc}
     \includegraphics[height=5.5cm]{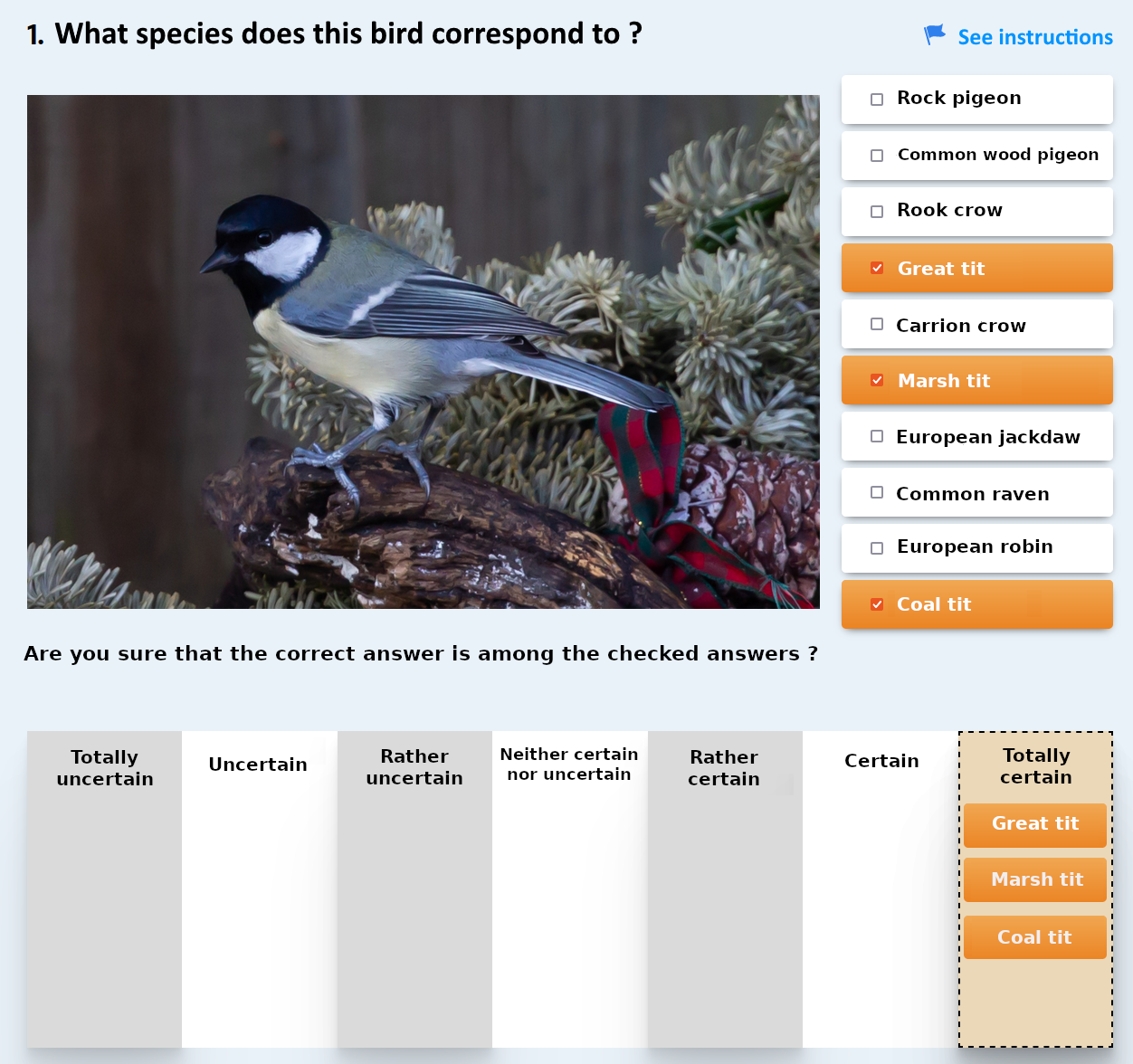} \quad & \includegraphics[height=5.5cm]{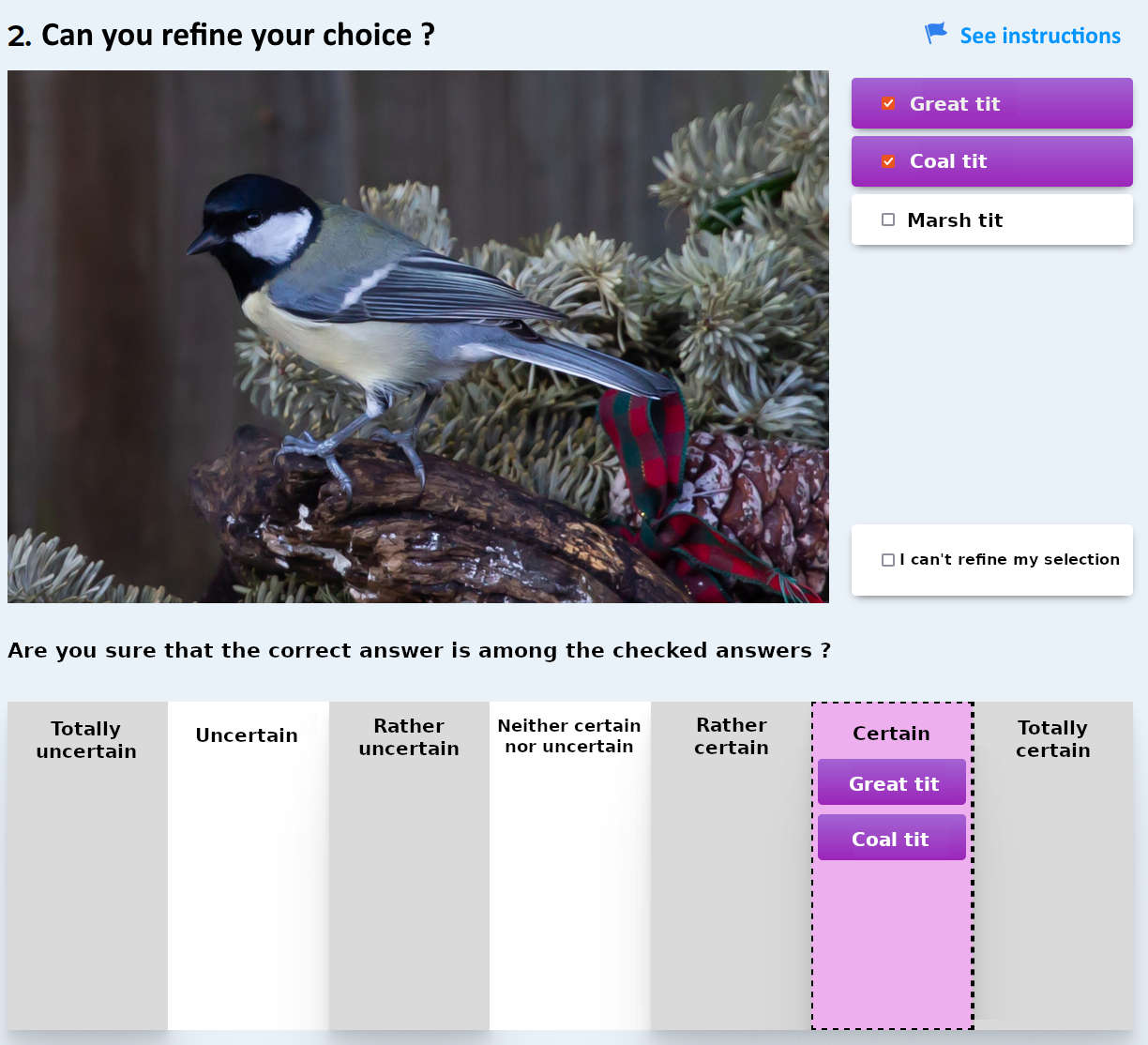} \\
    (a) iterative step 1 & (b) iterative step 2
\end{tabular}
    \caption{Interfaces used for the campaigns 10\_birds iterative and machine learning}
    \label{fig:interfaces_10_dyn}
\end{figure}


{\bf 10\_birds\_iterative}
This campaign is called iterative because the contributor is asked to expand or refine the contribution they have entered. 
To do so, in a first step the contributor answer the question as shown figure \ref{fig:interfaces_10_dyn}(a) and then:
\begin{itemize}
    \item If he is precise but not ``totally certain'' of his answer, he is offered to expand his selection if he feels the need.
    In this case, the first selected answer is kept in step 2 and he can complete it by selecting new names.
    \item If he is imprecise in his contribution, he is asked in a second step if he is able to restrict his choice of answer while giving his new certainty as in the example figure \ref{fig:interfaces_10_dyn}(b). 
    When he is offered to restrict his selection, only the previously chosen answer elements are proposed again.
\end{itemize} 
These interactions with the contributor increased the number of responses collected, and therefore the time spent soliciting the contributor. 


{\bf 10\_birds\_machine\_learning}
For classification problems, a larger number of observations is often required. This campaign provides imperfectly labeled observations that can be used for classification problems and more generally in machine learning. 
This is a similar campaign to 10\_birds\_iterative with more photos per class and fewer responses per photo. A number of 20 photos per bird species is used instead of 5. A total of 200 photos separated into 10 species are then labelled. Contributors are only invited to give 20 answers, 2 randomly selected per species of bird. No attention questions are asked in this campaign.

Each crowdsourcing campaign required a crowd of 50 contributors.
As with the other campaigns, a contributor who has participated in one experiment cannot participate in another.
For each of the ten bird that make up the proposed answer set, a contributor is presented with 5 photos of a bird, so that he answers 50 questions.
Thus 2500 data are collected for the experiments 10\_birds\_precise, with precise answers, and 10\_birds\_imprecise, for which the contributor can choose up to five answers.
There is 2990 data collected for 10\_birds\_iterative because for this experiment, as there are 2550 first step answers and 440 second step answers.
Finally 1515 data are collected for 10\_birds\_machine\_learning, with 1040 first step answers and 475 second step answers. 

\begin{table}[t]
    \centering
    \caption{Summary of the crowdsourcing campaigns conducted, the certainty is asked for all the answers}
    \begin{tabular}{|l|c|c|c|}
        \hline
        Campaign & Answers & Crowd size & Number of answer \\
        \hline
        multi\_birds\_precise & Precise & 100 & 5000 \\
        multi\_birds\_imprecise & Imprecise & 100 &  5000 \\
        10\_birds\_precise & Precise & 50 & 2500\\
        10\_birds\_imprecise & Imprecise & 50 & 2500\\
        10\_birds\_iterative & Imprecise & 51 & 2990\\
        10\_birds\_machine\_learning & Imprecise & 52 & 1515\\
        \hline
    \end{tabular}
    \label{tab:data_oiseaux}
\end{table}

All crowdsourcing campaigns are summarized in Table \ref{tab:data_oiseaux} which includes the name of the campaign, whether the data collected is accurate or inaccurate, the number of contributors and the total number of contributions collected.

\section{Details of the data sets} \label{section:data}

The datasets are made available to the community on the INRIA git account https://gitlab.inria.fr/cthierry/imprecise\_uncertain\_dataset.
The repository is \linebreak structured as follows: a folder is associated with each crowdsourcing campaign presented above, there is also a csv file named ``answers\_multi\_birds.csv''.
This file is used as the answer set propose to the contributor of the multi\_birds campaign. 
It includes the following variable:
\begin{itemize}
    \item photo: the number of the bird photo display by the interface
    \item goodanswer: the true bird name of the photo
    \item answ1, answ2, answ3, answ4: other bird names propose as answer elements to the contributor in addition of the true bird name
    \item difficulty: an hypothesis of the difficulty of the question according to the author. Values range from 1 (easy question) to 5 (difficult question). In fact the difficulty observed is not correlated to those one supposed by the author.
\end{itemize}
Thus, it is possible to use the file ``answers\_multi\_birds.csv'' to construct a frame of discernment $\Omega_q = \{goodanswer,answ1,answ2,answ3,answ4 \}$ for each question $q$ of the multi\_birds campaigns. In the files named after the crowdsourcing campaigns there are several files providing different information.

{\bf Data} 
This csv file includes the contribution from the crowd for the bird annotation.
The contribution includes therefore a selection of bird names and a certainty associated.
The file includes:
\begin{itemize}
    \item log\_id: indicates the line of the file
    \item user: unique user ID for a contributor
    \item currenttrial: number of the question asked to the contributor
    \item img: number of the photos shown to the contributor
    \item goodanswer: the true name of the bird to identify
    \item answer: the set of bird names selected by the contributor (a unique answer for the precise crowdsourcing campaign)
    \item answerhistory: set of values checked/unchecked by the contributor to answer the question 
    \item isgoodanswer: boolean indicated if the true bird name is include in the answer set given by the contributor
    \item certitude: certainty given by the contributor to express his confidence in his bird names selection
    \item certitudehistory: history of the certainty values for the answer
    \item timestamp: time recorded by the interface
    \item time: time of the contributor's answer to the question
\end{itemize}
Some variables such as user are common to several files. 
For all the files, the contributor certainty ranges from 1 (totally uncertain) to 7 (totally certain).

For the campaign multi\_birds\_imprecise 57.04$\%$  of the data includes in the csv file are imprecise.
And for the 10\_birds campaigns we have the following results: imprecise 55.64$\%$, iterative 45.32$\%$  and machine\_learning 58.22$\%$.
Contributors have made good use of the opportunity to be imprecise when possible. On average, when contributors are imprecise they choose two answers.

{\bf Attention}
In crowdsourcing campaigns, attention questions are asked to the contributor in order to ensure its seriousness. 
This csv file includes the answers to the three attention questions.
These questions consist in asking the contributor a previous question in order to get him to give exactly the same contribution. 
The variables included in the file are the following:
\begin{itemize}
    \item log\_id, user, currenttrial, answer, certitude, timestamp, time
    \item attention\_answer: contributor's answer to the attention question 
    \item answerhistory: set of values checked/unchecked by the contributor to answer the attention question 
    \item certainty: certainty of the first selected answer set
    \item certaintyhistory: history of checked/unchecked certainty values
    \item issamecertitude: boolean indicating if the certainty given to the attention question is identical to the certainty given to the initial question
    \item issameanswer: boolean indicating if the set of bird names selected at the attention question is identical to the set initially selected
\end{itemize}

{\bf Event}
This file records the principal events of the platform named event\_type: the connection to the platform (start), the beginning of the crowdsourcing campaign (start\_xp), the ending question (questions) and the end of the campaign (finish).
It is possible that some contributors have started a crowdsourcing campaign without having finished it, we have only a part of the answers for them. To sort out the contributors (users) to be selected we recommend using the event.csv or question.csv files described below to select the data of users who have reached the final question phase and/or the finish event. When we talk about the number of responses, it is only for contributors who have completed the entire campaign.
This file also includes the variables log\_id, user and time which gives the date and time when the event took place.

{\bf Queries}
At the end of the different crowdsourcing campaigns, a questionnaire is sent to the contributors to get their feedback. This questionnaire varies between campaigns. These files include the answers at the end of the campaign.

\begin{table}[t]
    \centering
    \caption{Number of precise and then imprecise contributions ($|X_1|=1$ and $|X_2|>1$) and imprecise then less imprecise ($|X_1|>|X_2|$).}
    \begin{tabular}{|c|c|c|c|}
        \hline
        \multirow{2}{*}{Campaigns} & Size of the  & \multirow{2}{*}{Data subset} & Size of the  \\
                                   &  dataset    &                              & of the data subset\\
        \hline
        \multirow{2}{*}{10\_birds\_iterative}  & \multirow{2}{*}{440} &  $|X_1|=1$ and $|X_2|>1$ & 88\\
        \cline{3-4}
         & & $|X_1|>1$ and $|X_1|>|X_2|$ & 352\\
        \hline
        \multirow{2}{*}{10\_birds\_machine\_learning}  & \multirow{2}{*}{475} &  $|X_1|=1$ and $|X_2|>1$ &  57\\
        \cline{3-4}
         & & $|X_1|>1$ and $|X_1|>|X_2|$ & 418\\
        \hline
    \end{tabular}
    \label{tab:nb_rep_w3}
\end{table}

{\bf Iteration}
This csv file is present in the 10\_birds\_machine\_learning and \linebreak 10\_birds\_iterative folders because it includes the contributors' answers when they expand or specify their answer in the second stage of questioning of these campaigns.
The next values are included into the file:
\begin{itemize}
    \item log\_id, user
    \item trial: equals to currenttrial value of the same contributor (user) in data.csv 
    \item new\_answer: new answer given by the contributor
    \item new\_certitude: new certainty given by the contributor
    \item cant\_answer: boolean that takes the value 1 if the contributor cannot modify (refine or enlarge) his answer
    \item isImprecis: boolean which takes the value 1 if in his first answer the contributor is imprecise ({\em i.e.} he chooses several bird names)
    \item aHistory: equivalent to answerhistory
    \item cHistory: equivalent to certaintyhistory
\end{itemize}
The file for the campaign 10\_birds\_iterative includes 1527 rows but for the majority of them, the contributor did not modify his answer (cant\_answer=1).
Indeed, for this campaign only 440 responses were modified which represents 17$\%$ of the first step answers. More contributors edited their answer for the 10\_birds\_machine\_learning campaign, 475 responses were modified i.e. 46$\%$ of the dataset.
Thanks to the joint use of this file and data.csv it is possible to build 440 consonant mass functions.

For campaigns with iteration we call $X_1$ the first set of names given for a photo. The values of $X_1$ (answer) are present in the data.csv file with the associated certainty.
When the contributor is proposed to modify his contribution, the new name selection $X_2$ (new\_answer) and the new certainty (new\_certitude) are registered in the iteration.csv file. 

For the campaign 10\_birds\_iterative there are 461 entries in the data.csv file for which the contributor first selected a single answer $|X_1|=1$, and then was offered to expand his selection so that $|X_2|>1$.
Of these 461 times when the contributor is offered to be imprecise, there are only 88 times when a second answer $X_2$ is given.
Similarly, there are a total of 1066 times when the contributor fills in an imprecise answer, $|X_1|>1$, and is offered to narrow his selection so that $|X_1| > |X_2|$, a total of 352 contributions report a change in answer.

During the 10\_birds\_machine\_learning campaign contributors also tend to give a second answer more precise than the first one, with 418 responses on 475 iterations, against 57 precise answers at firt step and less precise at second step. 
Furthermore, 389 responses were listed as totally certain and 186 were listed as totally certain and precise. Among those contributors who were certain and precise, 91$\%$ hold the real answer. Of all the answers, 33 were listed as inconsistent, which means that a contributor gave an answer including different bird families, these 33 answers were generated by 15 different contributors.

The following section presents examples of modeling with mass functions.

\section{Belief functions from the data}
\label{section:model}

We propose a data modeling by simple support and consonant mass functions.

{\bf Simple support mass function}
This function can be computed for the answer values and certainty of the six campaigns.
Given a question $q$, the set of answers associated to $q$ compose the framework of discernment $\Omega_q=\{r_1,\ldots,r_K\}$.
The question being closed, we consider the closed world. 
The contributor $c$ answers the question $q$ by the contribution $X \in 2^{\Omega_q}$, which can be imprecise, and to which he associates a certainty of value $certainty \in [1,7]$ which is transformed into a mass $\omega_{cq} \in [0,1]$ according to the equation:
\begin{equation}
    \omega_{cq} = \frac{certainty-1}{certainty_{max}-1}
    \label{eq:mass}
\end{equation}
For the crowdsourcing campaign introduce in this paper $certainty_{max}=7$.
A mass function with simple support ($X^{\omega_{cq}}$) can be obtained from the contribution:
\begin{equation} 
    \left \{
    \begin{array}{l}
        \displaystyle m_{cq}^{\Omega_q}(X) = \omega_{cq}   \mbox{ with } X \in 2^{\Omega_q} \setminus \Omega_q \\
        \displaystyle m_{cq}^{\Omega_q}(\Omega_q) = 1 - \omega_{cq} 
    \end{array}
    \right.
    \label{eq:X^w}
\end{equation}

{\bf Consonant mass function}
During the 10\_birds\_machine\_learning and \linebreak 10\_birds\_iterative campaigns, the same question $q$ can be asked twice to the contributor $c$ who can then enlarge or specify his first answer $X_1$ by a second answer $X_2$ if he wishes.
Let $\Omega_q$ be the set of proposed answers and $X_1, X_2 \in 2^\Omega_q$.
If the first answer of the contributor $X_1$ is precise and he widens his second answer $X_2$ then $X_1 \subset X_2$, and conversely if $X_1$ is more imprecise than $X_2$ then $X_2 \subset X_1$.
At the time of his first selection $X_1$, the contributor informs a degree of certainty of numerical value $\omega_{cq1} \in [0,1]$ compute thanks to equation~\eqref{eq:mass}. 
If he chooses to fill in a second answer $X_2$ he must indicate his new certainty whose numerical value is noted $\omega_{cq2} \in [0,1]$.
If the contributor is not asked to modify his selection or if he does not wish to do so, the contribution is modeled by a simple support mass function.
In the case where the contributor changes its response $X_1$ to the response $X_2$, with $X_1 \subset X_2$, then the contribution can be modeled by a consonant mass function:
\begin{eqnarray}
 \left \{
 \begin{array}{l}
  \displaystyle m_{cq}^{\Omega_q}(X_1) = {\delta_1} * \omega_{cq1} \\
  \displaystyle m_{cq}^{\Omega_q}(X_2) = {\delta_2} *{\omega_{cq2}} \\
  \displaystyle m_{cq}^{\Omega_q}(\Omega) = 1 - {\delta_1} * \omega_{cq1} - {\delta_2} *{\omega_{cq2}}
 \end{array}
 \right.
 \label{eq:mcon}
\end{eqnarray}
In equation~\eqref{eq:mcon}, the coefficients $\delta_1$ and $\delta_2$ ensure that the mass function belongs to the interval $[0,1]$, thus: ${\delta_1} + {\delta_2} = 1$.
If we want to give more importance to the first contribution $X_1$ rather than to the second contribution $X_2$ then we must choose $\delta_1$ such that $\delta_1>\delta_2$. 
Another way to combine the two mass functions from the two iterative responses is to use a combination rule that does not require the assumption of source independence.

We have proposed a modeling of some data by belief functions but it is possible to go further by using them for example to estimate the expertise of the contributor as do \cite{brjab16}. 
The data can also be used to compare a probabilistic approach to belief functions \cite{koulougli16}.

\section{Conclusion} \label{section:conclusion}

This paper presents some real credal datasets created through crowdsourcing campaigns for bird photo annotation. 
To constitute these datasets six crowdsourcing campaigns have been realized. 
In these six campaigns, the contributor is asked to give his certainty in his answer. 
For two campaigns the contributor is forced to choose a single bird name as an answer, these data are therefore precise and potentially uncertain. 
For the other four campaigns the contributor had the possibility to be imprecise in case of hesitation on his answer, these data are imprecise and/or uncertain. 
For these six crowdsourcing campaigns it is possible to model the contributions by simple support mass functions. 
Finally, for two of the four imprecise campaigns, the contributor is asked to modify the answer already given by clarifying or expanding it. 
Thanks to these two campaigns it is possible to model the contributions by consonant mass functions.


\bibliography{ref/belief}

\end{document}